\documentclass{ws-ijmpe}

\begin{document}

\markboth{G.Ramachandran, Swarnamala Sirsi and S. P. Shilpashree}{Photodisintegration of aligned deuterons at astrophysical energies using 
linearly polarized photons}

\catchline{}{}{}{}{}

\title{{\bf PHOTODISINTEGRATION OF ALIGNED DEUTERONS AT ASTROPHYSICAL ENERGIES USING LINEARLY POLARIZED PHOTONS }}

\author{G.Ramachandran}
\address{G.V.K. Academy, Bangalore, India\\
gwrvrm@yahoo.com}

\author{Swarnamala Sirsi}

\address{Department of Physics, Yuvaraja's College, University of Mysore, Mysore, India\\
sm\_sirsi@yahoo.co.in}

\author{S. P. Shilpashree \footnote{Also at Department of Physics, Yuvaraja's College, University of Mysore, Mysore, India and G.V.K. 
Academy, Bangalore, India}}
\address{Department of Physics, K.S.School of Engineering and Management, K.S.G.I, Bangalore, India\\
shilpashreesp@gmail.com}

\maketitle

\begin{history}
\received{(received date)}
\revised{(revised date)}
\end{history}

\begin{abstract}
Following the model independent approach to deuteron photodisintegration with linearly polarized $\gamma-$rays, we show that the measurements 
of the tensor analyzing powers on aligned deuterons along with the differential cross section involve five different linear combinations of 
the isovector $E1^j_v; j=0,1,2$ amplitudes interfering with the isoscalar $M1_s$ and $E2_s$ amplitudes. This is of current interest in view 
of the recent experimental finding \cite{blackston1} that the three $E1^j_v$ amplitudes are distinct and also the reported experimental 
observation \cite{sawatzky} on the front-back (polar angle) asymmetry in the differential cross section.
\end{abstract}
\noindent \\
\section{Introduction}

The earliest studies on deuteron photodisintegration and its inverse reaction viz., n-p fusion go back to more than seven decades 
\cite{chadwick}. These reactions are relavent to nucleosynthesis scenarios \cite{burbridge} of cosmological interest from the Big Bang to 
stellar evolution under various environments. The earliest estimates of the reaction rates by Fowler, Caughlan and Zimmermann \cite{fowler} 
used the theoretical calculations \cite{bethe} of deuteron photodisintegration normalized to the then available thermal neutron radiative 
capture cross section measurements \cite{hughes}. Since the deutron is loosely bound, it is destroyed in stellar environments even before the 
stars reach the main sequence and as such it has been observed \cite{smith} that `the ratio of the primordial abundance of deuterium  to that 
observed today could be anywhere between 1 and 50'. Although the measurements \cite{burles} of deuterium abundance in high-red-shift hydrogen 
clouds have improved the scenario, Burles et al \cite{bntt} have highlighted the need for the precise knowledge of the reaction rates for 
$d+\gamma\rightleftharpoons n+p$ at astrophysically relevant energies which correspond to the c. m. neutron energy range 25 to $200$ keV. \\
\\
While studies on photodisintegration of the deuteron are well documented \cite{arenhovel,hara,moreh} at lab photon energies $E^L_\gamma \geq 
2.33$ MeV and a large number of measurements exist for the n-p fusion cross section at thermal neutron energies which correspond to 
$E^L_n=10^{-5}$ keV  \cite{cox,cokinos}, there are only two measurements  at astrophysically relevant energies by Suzuki et al., \cite{7} at 
lab neutron energies  $E^L_n =20$, 40 and 60 keV and by Nagai et al., \cite{nagai} at $E^L_n=550$ keV as shown in Table 1. It was known quite 
early that the thermal neutron cross section is dominated by the isovector magnetic dipole amplitude denoted as $M1_v$. Breit and Rustgi 
\cite{breit} were the first to propose a polarized target-beam
experiment to look for an isoscalar magnetic dipole amplitude denoted as $M1_s$,  
in view of the then existing 10$\%$ discrepancy between theory 
and experiment. Although the meson exchange current contributions (MEC) 
suggested by Riska and Brown \cite{riska} explained this discrepancy in the cross section with surprising accuracy, 
the measured values for analyzing powers in $p(\vec n, \gamma)d$ as 
well as for neutron polarization in
photodisintegration of the deuteron were both found to differ 
\cite{holt,soderstrum,stephenson} from theoretical 
calculations which included MEC effects. In fact, it has also
been pointed out by Rustgi, Vyas and Chopra \cite{rustgi1} that there is an unambiguous 
disagreement between theory and experiment when a comparison  is made with the 
data at a photon energy of 2.75 MeV. They pointed out moreover, that the inclusion of
the two body effects widens this disagreement.\\
\\ Several theoretical studies \cite{partovi,cambi} have been reported  on $d+\gamma\rightleftharpoons n+p$
based on a variety of deuteron models, including those at quark level. Some of these calculations take into account meson exchange and other 
currents.
Nagai et al., \cite{nagai} have quoted  the theoretical results of Sato et al., \cite{sato} which included MEC, isobar currents and pair 
currents. These theoretical results 
show that the $M1_v$ strength decreases
sharply with increasing energy, while $E1_v$ i.e., the isovector electric dipole amplitude starts appearing in the energy range of interest 
to astrophysics and  increases with energy there after. 
Moreover, the theoretical calculations quoted by Nagai et. al., \cite{nagai} show that the $E1_v$ and $M1_v$ strengths are approximately 
equal at around c.m neutron energy $E_n$ of order $500$ keV.
The relative strengths of $M1_v$ and $E1_v$ have also been discussed in several studies using effective field theory \cite{ando}. While the 
role of isoscalar $M1_s$ have been discussed by Chen et al., \cite{11} and Park et al., \cite{park} using different versions of effective 
field theory. The predictions of these two versions led to an experimental measurement of the $\gamma$ anisotropy by Muller et al., 
\cite{muller}, which however could not decide in favor of either of the approaches. The calculations of $E2_s$ contributions have also been 
discussed by Chen et al., \cite{11} and Hadjimichael et al., \cite{had}. Although the $E1_v$ transition dominates photodisintegration of 
deuteron, the $E2_s$ contributions, though small, has been considered to be still significant \cite{schivilla}. 
\\
An observable
which is sensitive to the presence of isoscalar $M1_s$ and $E2_s$ transitions from the
triplet $S$-state is the circular polarization of the emitted radiation with
initially polarized neutrons. The first measurement \cite{22} to detect the
presence of isoscalar amplitudes was not quite encouraging but a subsequent
measurement \cite{23} yielded a value $P_\gamma= -(2.29 \pm 0.9)\times 10^{-3}$.
An attempt \cite{24} to explain the large measured value by introducing a
six quark admixture in the deuteron wave function led, however, to a 
disagreement with the precise deuteron magnetic moment. Later calculations
 \cite{25} in the zero range approximation and the wavefunction for a Reid soft
core potential led to a theoretical prediction $P_\gamma $ of the order of 
$-1.1 \times 10^{-3}$ with an estimated accuracy of 25$\%$. The latest measured
value \cite{26} of $P_\gamma =-(1.5 \pm 0.3)\times 10^{-3}$ is in 
reasonable agreement with the theoretical calculation \cite{25}. The 
importance of measuring the photon polarization with initially polarized
neutrons incident on a polarized proton target has been pointed out \cite{27}.
When the initial preparation of the neutron and proton polarizations
$P(n)$ and $P(p)$ are such that they are either opposite to each
other or orthogonal to each other, the interference of the small
isoscalar amplitudes with the large isovector amplitude could substantially
contribute to the observable photon polarization. Attention has also been focussed on the spin response of the deuteron, i.e., the asymmetry 
of photoabsorption with respect to parallel and antiparallel spins of photon and deuteron and the Gerasimov-Drell-Hearn (GDH) sum rule 
\cite{gdh1,gdh2} by Arenhovel et al., \cite{aren1}, who have also considered other polarization observables in electromagnetic deuteron 
break-up \cite{aren2} and compare their formalism with that of Dmitrasinovic and Gross \cite{gross}. 
We might also mention an experimental measurement of recoil proton polarization in electrodisintegration of deuteron by polarized electrons 
\cite{barkhuff} and a theoretical discussion of  photodisintegration of polarized deuterons by unpolarized photons \cite{sps1} using the 
model independent formalism \cite{sps}.
\\
\\
\begin{table}[t]
\caption{Measured cross-sections}
\begin{tabular}{cccccc}
\hline
&   for $p(n,\gamma)d$ &    &   & for $d(\gamma,n)p$\\
\hline
Energy  & Cross-section & Reference & Energy  & Cross-section & Reference\\
$E^L_n$ keV   & mb &            & $E^L_\gamma$ MeV   & mb &\\
\hline
$10^{-5}$    & $334.2 \pm 0.5$ & ~\cite{cox}  &$2.33$ &   $0.683 \pm 0.053 \pm 0.042$ & ~\cite{hara}\\
$10^{-5}$   & $332.6 \pm 0.7$ &  ~\cite{cokinos} & $2.52$ &   $0.983 \pm 0.039 \pm 0.061$ & ~\cite{hara}\\
$20$ &    $(318 \pm 2) \times 10^{-3}$ & ~\cite{7} & $2.62$ &   $1.300 \pm 0.029$ & ~\cite{arenhovel}\\
$40$ &   $(203 \pm 19) \times 10^{-3}$  & ~\cite{7} & $2.754$ &   $1.456 \pm 0.045$ & ~\cite{moreh}\\
$64$ &   $(151 \pm 7) \times 10^{-3}$ & ~\cite{7} & $2.76$ &   $1.474 \pm 0.032$ & ~\cite{arenhovel} \\
$550$ &   $(35.2 \pm 2.4) \times 10^{-3}$ & ~\cite{nagai}  & $2.79$ &   $1.47 \pm 0.03 \pm 0.09$ & ~\cite{hara}\\
\hline
\end{tabular}
  \end{table}
As the Big Bang Nucleosynthesis(BBN) entered the precision era \cite{schramm}, an accurate estimation of 
the primordial 
abundance of deuterium  became important and it has been referred to as the Cosmic 
Baryometer. The work of Burles et al., \cite{burles} catalyzed the first experimental study \cite{schreiber} on the analyzing power 
$\Sigma(\theta)$ in deuteron photodisintegration at $E^\gamma_L=3.58$ MeV and $\theta=150^o$, using 100 $\%$ linearly polarized photons from  
the High Intensity Gamma-ray source (HIGS) at the Duke Free Electron Laser Laboratory. It was followed by further studies 
\cite{tornow1,tornow2,sawatzky,blackston,ahmed,blackston1}. These measurements  have been analyzed, using a theoretical formalism 
\cite{weller} where the $M1_v$ and $E1_v$ multipole contributions (which are dominant at energies of thermal neutron capture and deuteron 
photodisintegration respectively) were calculated separately. 
The model independent theoretical formalism \cite{sps} for deuteron photodisintegration with linearly polarized photons showed that the 
differential cross section in $d(\vec \gamma,n)p$ contains  a term representing the interference between the $E1_v$ and $M1_s$ amplitudes, 
which is non zero if the three $E1_v$ amplitudes $E1_v^j$, $j=0,1,2$ are unequal. In a recent publication, Blackston et al., 
\cite{blackston1}  have reported the first experimental observation of the splitting of the three $E1_v$ amplitudes. 
\\
\\ Therefore, the purpose of the present paper is to employ the model independent theoretical approach \cite{sps} to carry further the study 
of the contributions of isoscalar $M1_s$ and $E2_s$ amplitudes in photodisintegration of deuterons by linearly polarized photons using 
aligned deuteron targets. 
\section{Aligned deuteron}
The deuteron is a spin-1 nucleus, which gets polarized in different ways in different external environments. For example, an oriented 
\cite{blino} deuteron target can be produced under the influence of an external uniform magnetic field, whose direction determines the axis 
of orientation. On the other hand, when the deuteron is exposed to external electric quadrupole fields generated by surrounding electrons in 
crystal lattice sites \cite{edmonds}, the spin of the deuteron is aligned.  States of polarization of spin-1 nuclei exist, which are more 
complex and are in fact multiaxial \cite{mvn}. The state of polarization of a spin-1 nucleus is, in general, specified by the spin density 
matrix
\begin{equation}
\rho= \frac {\textrm{Tr} \rho}{3}\left\{
    \begin{array}{ccc}
     1+\frac{3}{2} t^1_0+\frac{1}{\sqrt 2} t^2_0\quad & \frac{3}{2} (t^1_{-1}+t^2_{-1}) &\quad \sqrt{3} t^2_{-2}\\
      -\frac{3}{2} t^1_1+t^2_1 \quad & 1-\sqrt{2} t^2_0 \quad & \frac{3}{2} (t^1_{-1}+t^2_{-1})\\
	  \sqrt{3} t^2_{2}\quad & -\frac{3}{2} (t^1_{1}+t^2_{1}) \quad & 1-\frac{3}{2} t^1_0+\frac{1}{\sqrt 2} t^2_0
	       \end{array} \right\}
\end{equation}
which is hermitian and is parametrised in terms of the 
 Fano \cite{fano}  statistical tensors $t^k_q, q=k,k-1,...,-k$ with $k=1$ referring to its vector polarization and $k=2$ referring to its 
tensor polarization. The normalization used for the $t^k_q$ in eq. (1) follows \cite{mvn,madison,ksm,devi}. 
The rows and columns of the above matrix are labeled by the states $|1 m>$ of the deuteron spin-1 with magnetic quantum numbers $m=+1, 0, -1$ 
in that order w.r.t any chosen quantization axis, $\hat Q$ referred to usually as the Z-axis. The target is said to be unpolarized if all the 
$t^k_q=0$. Defining vector polarization through
\begin{eqnarray}
t^1_0= \sqrt{\frac {3}{2}} P_z\qquad
t^1_{\pm 1}=\mp \frac{1}{2} \sqrt{3}(P_x+iP_y)
\end{eqnarray}
the target is said to be purely vector polarized if the vector polarization $\vec P \neq 0$  but $t^2_q=0$. Defining the tensor polarization 
through a traceless symmetric 2nd rank cartesian tensor $T_{\alpha\beta}; \alpha,\beta=x,y,z$ given by
\begin{equation}
t^2_0=\frac{1}{\sqrt 2}(T_{zz}); \qquad t^2_{\pm 1}= \mp \frac {1}{\sqrt 3} (T_{xz}\pm i T_{yz}) \qquad t^2_{\pm 2}= \frac {1}{2 \sqrt 
3}(T_{xx}-T_{yy}\pm 2 i T_{xy}),
\end{equation}
a spin 1 nucleus like the deuteron is said to be aligned, if its vector polarization is zero and tensor polarization is non-zero.
\\
\\
 Although the energy levels of spin 1 nuclei like deuteron (with  non-zero electric quadrupole moment) have been studied in external electric 
quadrupole fields \cite{lucken}, Ramachandran, Ravishankar, Sandhya and Swarnamala Sirsi \cite{sandhya} are probably the earliest to draw 
attention to the non-orientedness of the nuclear polarization in such an environment when the asymmetry  parameter $\eta$  of the external 
electric quadrupole field with strength $A$ is non-zero. Precise estimates of the parameters $A$ and $\eta$ when they are embedded in various 
compounds and at different sites are available \cite{edmonds,lucken}. The estimates of $t^2_q$ are also available \cite{swarna} at 
temperature of order mK.
With the present day technology, in the case of ultracold atoms in optical lattice, the lowest temperatures reached are of the order of nK 
\cite{temp1} and pK \cite{temp2}. It could therefore be feasible to prepare aligned deuteron targets with higher values of tensor 
polarization.\\
\\
 An  external electric quadrupole field tensor  $V_{\alpha\beta}$
  is represented by $V_{XX}, V_{YY}$ and $V_{ZZ}$ in the Principal Axis Frame (PAF) \cite{swarna}, where $V_{\alpha\beta} \propto 
{\delta_\alpha\beta}$. When a spin-1 nucleus with a non-zero nuclear electric quadrupole moment $Q$ is exposed to such an external electric 
quadrupole field, the resulting polarization is describable in terms of the Fano statistical tensors $t^k_q$ where $t^1_q=t^2_{\pm 
1}=Im(t^2_2)=0$, which defines the Principal Axes of Alignment Frame (PAAF) \cite{ksm} which coincides with PAF in this case when no other 
external fields like magnetic field are present. The interaction Hamiltonian is well known and is of the form
\begin{equation}
H_{int}=A[{(3 J_Z^2-J^2)+\eta(J_X^2-J_Y^2)}],
\end{equation}
where $J_X, J_Y$ and $J_Z$ are the cartesian components of the nuclear spin $\vec J$ with $J^2=\vec J \cdot \vec J$ and $A$ is proportional 
to the nuclear electric quadrupole moment $Q$ through
\begin{equation}
A={1 \over 4}Q V_{ZZ}, \qquad  \eta={V_{XX}-V_{YY} \over V_{ZZ}}; \quad |V_{ZZ}|\geq |V_{YY}|\geq |V_{XX}|.
\end{equation}
The resulting eigen states $|X>,|Y>,|Z>$ with $A(1 \pm \eta)$ and $-2A$ energy eigen values respectively are given by 
\begin{eqnarray}
|X> = {1 \over \sqrt 2} {(|1 -1>-|1 1>)}; \quad E_X=A(1+\eta) \nonumber \\
|Y> = {i \over \sqrt 2} {(|1 -1>+|1 1>)}; \quad E_Y=A(1-\eta) \nonumber \\
|Z> = |1 0>; \quad E_Z=-2A,
\end{eqnarray}
in terms of magnetic substates $|1m>$ defined with respect to the Z-axis of the PAAF, where $t^2_{\pm 1}=0$ \\
\\
The state of the aligned deuteron is characterized by two axes \cite{mvn} which are along $\hat {n_1}=(\theta_1,\phi_1)$ and $\hat 
{n_2}=(\theta_2,\phi_2)$ in PAAF. The $\pm \hat {n_1}$ and $\pm \hat {n_2}$  are determinable by setting $t^2_2=0$, which leads to a 
quadratic equation having in general two solutions. In the particular case where $\hat {n_1}$ and $\hat {n_2}$ are collinear, the polarized 
spin-1 system is oriented. Otherwise it is said to be nonoriented. Some typical cases \cite{swarna} are illustrated in Table 2.
\begin{table}[t]
\caption{Polarization parameters of aligned deuterons at temperature T =0.1 mK}
\begin{tabular}{cccccc}
\hline
Compound  & A  & $-\eta$ & $t^2_0$ & $t^2_2$  \\
\hline
acetamide (ND2)   & 0.6242 & 0.4022  &-0.4584 &   -4.6791E-02 \\
chloroacetamide (ND2)  & 0.6260 & 0.3772 & -0.4601 & -4.3949E-02 \\
formamide (ND2)&  0.5752 & 0.3712 & -0.4223 &  -4.1353E-02 \\
phthalamide $(ND_2)_{II}$& 0.6457  & 0.3722 & -0.4749 &  -4.4035E-02 \\
L-asparagine (ND2) & 0.6377 & 0.3826 & -0.4688 & -4.4990E-02  \\
L-asparagine hydrate (ND2) & 0.6797 & 0.3272  & -0.5010 & -3.9644E-02 \\
\hline
\end{tabular}
  \end{table}
\section{Model Independent Theoretical Formalism}
Following \cite{sps} and using the same notations, the reaction matrix for $d+\gamma \to n+p$ with linearly polarized photons is 
\begin{equation}
{\bf M} = \sum_{s=0}^{1} \sum_{\lambda = |s-1|}^{s+1}
(S^\lambda(s,1) \cdot {\mathcal F}^\lambda(s)),
\end{equation}
where $S^\lambda_{\nu}(s,1)$ are irreducible tensor operators of rank
$\lambda$ in hadron spin space \cite{39} connecting the initial spin 1 state
of the deuteron with the final singlet and triplet states, $s=0,1$ of 
the $n-p$ system in the continuum. The 
irreducible tensor amplitudes, ${\mathcal F}^{\lambda}_\nu 
(s)$ are given by
\begin{equation}
{\mathcal F}^{\lambda}_{\nu}(s)=\sum_{\mu=\pm 1} {\mathcal F}^\lambda_\nu(s,\mu),
\end{equation}
where the four irreducible tensor amplitudes, ${\mathcal F}^\lambda_\nu(s,\mu)$ may be written, taking into consideration both the isoscalar 
amplitudes $M1_s$ and $E2_s$, in addition to the isovector $M1_v$ and the three $E1_v^j, j=0,1,2$ amplitudes, as
\begin{eqnarray}
{\mathcal F}^0_0(1,\mu)={1 \over 3} E1_v(0) f^0_0(1,1,\mu); \nonumber \\
{\mathcal F}^1_\nu(0,\mu) =
-i M1_v f^1_\nu (0,1,\mu)
 \nonumber \\
{\mathcal F}^1_\nu(1,\mu) = -{1 \over 6} E1_v(1) f^1_\nu(1,1,\mu) 
+ i M1_s f^1_\nu (0,1,\mu);
\nonumber \\
{\mathcal F}^2_\nu(1,\mu) = {1 \over 6}E1_v(2) f^2_\nu(1,1,\mu)
 +E2_s  f^2_\nu (0,2,\mu).
\end{eqnarray}
The $E1_v(\lambda=0,1,2)$ in the above equation are linear combinations of the $E1_v^j$ amplitudes with  $j=0,1,2$. They are explicitly given 
by
\begin{equation}
\left[
    \begin{array}{c}
     E1_v(0)\\
     E1_v(1) \\
     E1_v(2)
     \end{array}
     \right ]
= \left [
\begin{array}{ccc}
1 & 3 & 5 \\
2 & 3 & -5 \\
2 & -3 & 1
\end{array}
\right ]
\left [
\begin{array}{c}
E1_v^{j =0}\\
     E1_v^{j =1} \\
     E1_v^{j =2}
     \end{array}
     \right ].
\end{equation}
 and $f^\lambda_\nu (l,L,\mu)$ are given by
 \begin{eqnarray}
f^\lambda_\nu (l,L,\mu) = (2\pi)^{3/2}  (i\mu)^{\pi^+}
C(l, L, \lambda ; m_l, -\mu, \nu)
Y_{lm_l}(\theta,\phi),
\end{eqnarray}
which take care of the angular dependence and also the
dependence on photon circular polarization $\mu=\pm 1$.
\\
\\
The differential cross section with linearly polarized photons is then given by
\begin{eqnarray}\label{diff}
{d\sigma \over d\Omega} =
{2\pi^2 \over 6} [{a+b \sin^2\theta (1+\cos 2\phi)
-c\cos\theta}].
\end{eqnarray}
where
\begin{equation}
c= 4\sqrt 6\big[Re[(E1_v(1))M1_s^*]+Re[(E1_v(2))E2_s^*]\big].
\end{equation}
This result, viz., eq. (12), for the differential cross section with polarized photons bears comparison with the form for the differential 
cross section obtained by Rustgi et al., in their approximations $D$ and $E$ which take into account the $M1_s$ and $E2_s$ amplitudes. The 
$\cos \theta$ term is present in both these approximations. 
The presence of the term with coefficient $c$ can easily be detected by measuring the front-back (polar angle) asymmetry 
\begin{equation}
 {d\sigma \over d \Omega} (\theta)-{d\sigma \over d \Omega}(180-\theta)  = -{2\pi^2 \over 3}c\cos\theta
\end{equation}
In the abstract of his experimental study of $d(\vec \gamma,n)p$
at laboratory photon energies 3.5 to 10 MeV, Sawatzky\cite{sawatzky} has observed: ``Analysis of these data revealed a striking discrepancy 
with the prevailing low energy models of the Nucleon-Nucleon (N-N) interaction. The most prominent feature of the disagreement lies in an 
observed front-back (polar angle) asymmetry in the 
c.m. 
photoneutron yield that is not represented in the theory (e.g.,\cite{arenhovel,rupak}). The magnitude of this discrepancy rises as the 
$\gamma$-ray
energy falls towards the threshold". The last sentence in the above quote is particularly relevant for the discussion of the problem at 
astrophysical energies. 
\\
\\
The asymmetry represented by eq. (14) could also be caused by $f-$waves, i.e., with $l=3$ in the final state, but the electric multipoles in 
$f$ waves do not interfere with the $M1_s$ amplitude. 
Note also that $c$ reduces to zero, if all the three $E1_v^j; j=0,1,2$ amplitudes are equal: if so, the asymmetry represented by eq. (14) 
could possibly be caused by an $E1_v^{j=2}$ multipole amplitude leading to $l=3$ (which is different from three  $E1_v^j, j=0,1,2$ amplitudes 
leading to $l=1$) interfering with the $E2_s$ amplitude. But it was found recently \cite{blackston1} that that the three $E1_v^j$ amplitudes 
are not equal at 14 MeV and 16 MeV. Such an inequality is likely to be present even at low energies like 3.5 to 10 MeV employed in the 
experimental studies of Sawatzky \cite{sawatzky}. Therefore, it is worth while to carry out further studies on the inequality of the $E1_v^j$ 
amplitudes leading to $p$-waves at lower energies and their interference with the $M1_s$ and $E2_s$ amplitudes.
\section{Analyzing powers for deuteron disintegration by linearly polarized photons}
Let us therefore consider the differential cross section for photodisintegration of aligned deuterons by linearly polarized photons, which is 
given by 
\begin{eqnarray}
({d\sigma \over d\Omega})_{aligned} ={1 \over 6} Tr [{\bf M} \rho {\bf M^\dagger}],
\end{eqnarray}
where $\rho $ is given by equation (1) with non-zero $t^2_q$. This leads to 
\begin{equation}
({d\sigma \over d\Omega})_{aligned}  = {d\sigma \over d\Omega} {[1+({\mathcal A}^2 \cdot t^2)]}
\end{equation}
where the analyzing powers are given by
\begin{eqnarray}
{\mathcal A}^2_q = {1 \over{2 \sqrt 3}}\sum_{s,\lambda,\lambda'} (-1)^\lambda (2s+1) [\lambda][\lambda']W(1 2 s \lambda; 1 \lambda') 
({\mathcal F}^\lambda (s)
\otimes {\mathcal F}^{\dagger\lambda'} (s))^2_q \quad 
\end{eqnarray}
which are expressible in the form,
\begin{eqnarray}\label{a20}
A^2_0=  \sqrt 2 \pi^2[a_0+b_0 \sin^2 \theta +2c_0 \sin^2 \theta \cos 2\phi+d_0 \cos \theta]
\end{eqnarray}
where
\begin{eqnarray}\label{a}
a_0= \big[{4 \over 3} |M1_v|^2- 2|M1_s|^2 -2|E2_s|^2
- 3|E1_v(1)|^2 -3|E1_v(2)|^2 \nonumber \\ - 18 Re[(E1_v(2) E1_v^*(1)] 
-6Re(E2_s^{*} M1_s^{*}) \big],\qquad
\end{eqnarray}
\begin{eqnarray}\label{b}
 b_0 = 2 \big[3 |E1_v(1)|^2+ 4|E1_v(2)|^2
+ Re(2 E1_v^*(0)+9E1_v^*(1))E1_v(2)
 \big], 
\end{eqnarray}
\begin{eqnarray}\label{c}
 c_0 = \big[ {-3 \over 2}|E1_v(1)|^2-{1 \over 2}|E1_v(2)|^2
+ 2 Re (E1_v(2) E1_v^*(0))
 \big],
\end{eqnarray}
\begin{equation}\label{d}
d_0 = {4\over \sqrt 6}  [{3 \over 2} Re( E1_v(1) + 3E1_v(2))M1^{*}_s + {3 \over 2} Re(3 E1(1) + E1_v(2))E2_s^{*}].
\end{equation}
and
\begin{eqnarray}\label{a22}
A^2_2= 2 \sqrt 3 \pi^2[a_2+b_2 \sin^2 \theta +c_2 \sin^2 \theta e^{2i\phi}+d_2 \cos \theta].
\end{eqnarray}
\begin{eqnarray}\label{a}
a_2= \big[{2 \over 3} |M1_v|^2- |M1_s|^2 -|E1_s|^2
- {3 \over 2} |E1_v(1)|^2 - {3 \over 2}
|E1_v(2)|^2 \nonumber \\+3 Re(E1_v(2) E1_v^*(1)) 
+Re(E2_s M1_s^{*}) \big],\qquad
\end{eqnarray}
\begin{eqnarray}\label{b}
 b_2 = \big[ {3 \over 2} |E1_v(1)|^2+
{1 \over 2}|E1_v(2)|^2
- 2 Re(E1_v(2) E1_v^*(0)) 
 \big], 
\end{eqnarray}
\begin{eqnarray}\label{c}
 c_2 = \big[- |E1_v(2)|^2 - Re(2E1_v^*(0)+3E1_v^*(1))E1_v(2) 
 \big],
\end{eqnarray}
and
\begin{equation}\label{d}
d_2 = \sqrt{3 \over 2}  Re[( E1_v(2) - E1_v(1))(M1_s^{*}-E2_s^{*})\big].
\end{equation}
\section{Discussion}
It is important to note that the coefficients of $\cos \theta$ in the differential cross section ${d\sigma \over d\Omega}$ as well as in the 
analyzing powers $A^2_0$ and $A^2_2$(i.e., $c, d_0$ and $d_2$) involve interference between the three isovector $E1_v(\lambda)$ amplitudes 
given by eq. (10) and the isoscalar $M1_s$ and $E2_s$ amplitudes. Using eq. (10) and expressing
\begin{equation}
c = Re(\alpha M1_s^{*}+\beta E2_s^{*})
\end{equation}
\begin{equation}\label{d}
d_0 = Re(\alpha_0 M1_s^{*}+\beta_0 E2_s^{*})
\end{equation}
\begin{equation}\label{d}
d_2 = Re(\alpha_2 M1_s^{*}+\beta_2 E2_s^{*}),
\end{equation}
 where
\begin{equation}
\alpha=4\sqrt 6 (2 E1_v^{j=0} + 3 E1_v^{j=1} 
-5 E1_v^{j=2})
\end{equation}
\begin{equation}
\beta=4\sqrt 6 (2 E1_v^{j=0} - 3 E1_v^{j=1} 
+ E1_v^{j=2})
\end{equation}
\begin{equation}
\alpha_0= -4\sqrt 6  ( E1_v^{j=0} - 3E1_v^{j=1}+2E1_v^{j=2})
\end{equation}
\begin{equation}
\beta_0=-4\sqrt 6 (E1_v^{j=0} + 3E1_v^{j=0}-4E1_v^{j=2}).
\end{equation}
\begin{equation}
\alpha_2=-\beta_2=\sqrt{2 \over 3}  ( E1_v^{j=0} - E1_v^{j=2})
\end{equation}
in terms of three $E1_v^j, j=0,1,2$ amplitudes.
\\
\\
It is interesting  to note that the above equations (31) to (35) involve five different linear combinations of the three $E1^j_v$ amplitudes 
and as such facilitate their study individually at lower energies of interest to astrophysics. The experimental finding \cite{blackston1} at 
14 MeV and 16 MeV that all the three $E1^j_v$ amplitudes are not equal is encouraging. Since the possible role of the $M1_s$ and $E2_s$ 
amplitudes have been discussed by several authors using different theoretical formalisms over the last 50 years, we feel that it is desirable 
to carry out the measurements on the tensor analyzing powers $A^2_0$ and $A^2_2$ in photodisintegration of aligned deuterons in addition to 
the differential cross section ${d\sigma \over d\Omega}$. Such experimental studies are expected to clarify purely empirically the role of 
the isoscalar $M1_s$ and $E2_s$ amplitudes at the range of energies of interest to astrophysics.

\section{Acknowledgements}
We are thankful to Prof. Blaine Norum for making \cite{sawatzky,blackston} available to us and for helpful communications. We also thank 
Prof. Ando for drawing our attention to \cite{ando,cokinos}. One of us (SPS) is thankful to Principal and Management, K.S.S.E.M, Bangalore, 
for their encouragement.

\end{document}